# Entropic Stabilization of Proteins by TMAO

*Samuel S. Cho[†‡], Govardhan Reddy[†], John E. Straub[§], and D. Thirumalai[†*]*

[†]Department of Chemistry and Biochemistry and Biophysics Program, Institute for Physical Science and Technology, University of Maryland, College Park, Maryland 20742

[‡]Present Address: Departments of Physics and Computer Science, Wake Forest University, Winston-Salem, North Carolina, 27109

[§]Department of Chemistry, Boston University, Boston, Massachusetts 02215

thirum@umd.edu

TITLE RUNNING HEAD Molecular Mechanism for Protein-TMAO Stabilization

**Abstract**

The osmolyte trimethylamine N-oxide (TMAO) accumulates in the cell in response to osmotic stress and increases the thermodynamic stability of folded proteins. To understand the mechanism of TMAO induced stabilization of folded protein states, we systematically investigated the action of TMAO on several model dipeptides (Leucine, $L_2$, Serine, $S_2$, Glutamine, $Q_2$, Lysine, $K_2$, and Glycine, $G_2$) in order to elucidate the effect of residue-specific TMAO interactions on small fragments of solvent-exposed conformations of the denatured states of proteins. We find that TMAO preferentially hydrogen bonds with the exposed dipeptide backbone, but generally not with nonpolar or polar side chains. However, interactions with the positively charged Lys are substantially greater than with the backbone. The dipeptide $G_2$, is a useful model of pure amide backbone, interacts with TMAO by forming a hydrogen bond between the amide nitrogen and the oxygen in TMAO. In contrast, TMAO is depleted from the protein backbone in the hexapeptide $G_6$, which shows that the length of the polypeptide chain is relevant in aqueous TMAO solutions. These simulations lead to the hypothesis that TMAO-induced stabilization of proteins and peptides is a consequence of depletion of the solute from the protein surface provided intramolecular interactions are more favorable than those between TMAO and the backbone. To test our hypothesis we performed additional simulations of the action of TMAO on an intrinsically disordered $A\beta_{16-22}$ (KLVFFAE) monomer. In the absence of TMAO $A\beta_{16-22}$ is a disordered random coil. However, in aqueous TMAO solution $A\beta_{16-22}$ monomer samples compact conformations. A transition from random coil to $\alpha$-helical secondary structure is observed at high TMAO concentrations. Our work highlights the potential similarities between the action of TMAO on long polypeptide chains and entropic stabilization of proteins in a crowded environment due to excluded volume interactions. In this sense TMAO is a nano-crowding particle.

**Introduction**

Trimethylamine *N*-oxide (TMAO) is a naturally occurring osmolyte that accumulates in organisms to counteract the destabilizing effect of urea[1] on folded protein conformations. A number of experiments have shown that TMAO stabilizes proteins[2–4], but the precise molecular mechanism has not been firmly established[5–7]. The stabilization of proteins by TMAO can be qualitatively understood from the perspective of an entropic stabilization mechanism[8] introduced in the context of crowding effects on protein stability. Depletion of an osmolyte from the vicinity of proteins results in compact conformations, which stabilizes the native states[8–12]. On the other hand, if an osmolyte were to interact directly with the protein, as is the case with denaturants such as urea and guanidinium hydrochloride, the native basin of the protein would be destabilized[13–15]. These arguments, while rationalizing the different roles of protective and denaturing osmolytes, do not provide a molecular explanation of their actions.

The structure of TMAO (Fig. 1) suggests that there are two main types of intermolecular interactions that are possible between TMAO and proteins. The oxygen atom on TMAO ($O_T$) can act as a hydrogen bond acceptor. Three methyl groups in TMAO can participate in hydrophobic interactions with the sidechains of proteins. From transfer free energy calculations, it has been deduced that TMAO has no significant preference for hydrophobic moieties, but TMAO interacts with the backbone, as well as charged and polar sidechains[2]. It is necessary to extend such studies to systems in which chain connectivity and sequence effects are explicitly taken into account.

To dissect the molecular basis for the action of TMAO on peptides we simulated five dipeptides in explicit water in 1 M TMAO. The dipeptides are ideal model systems for the study of TMAO-protein interactions because, like unfolded proteins, they are solvent exposed, and hence can freely interact with the surrounding solvent molecules. Each dipeptide was composed of one of the following types of amino acids: leucine (nonpolar), serine (polar, hydroxyl group), glutamine (polar, amino group), and lysine (basic). In addition, we also studied conformation changes in diglycine ($G_2$) and hexaglycine

($G_6$), which has been recently investigated using MD simulations in aqueous TMAO solution[16]. Comparison of the conformational changes in $G_2$ and $G_6$ in TMAO leads to the hypothesis that as the polypeptide chain length inreases TMAO is expelled from the surface, which results in the collapse of the predominantly backbone construct. Consequently, $G_6$ adopts a conformation that maximizes the intra-peptide interactions. In order to further validate our hypthesis we performed all-atom MD simulations of $A\beta_{16-22}$ (KLVFFAE) monomer, which aggregates to form fibrils, in various TMAO concentrations. The $A\beta_{16-22}$ peptide, which consists of a short sequence of hydrophobic residues flanked by two oppositely charged residues is disordered and adopts a random coil conformtion that is devoid of secondary strucure. Remarkably, $A\beta_{16-22}$ becomes helical upon interaction with increasing concentrations of TMAO. Analysis of the conformations of $A\beta_{16-22}$ shows that TMAO is depleted from the surface of the backbone, which establishes that TMAO-induced transition from random coil to $\alpha$-helix is due the entropic stabilization mechanism. Thus, the stabilization of proteins by TMAO is akin to mechanism by which crowding particles stabilize proteins, which suggests TMAO can be treated as a nano-crowding particle.

**Methods**

We performed MD simulations using the NAMD program[17] and the CHARMM22[18] force field with the CMAP modification[19] for proteins and waters. In order to describe the interactions between the osmolyte and the polypeptide chains we used the TMAO force field parameters of Kast et al.[20]. We first simulated five dipeptides and one hexaglyine in order to dissect the interaction between TMAO and polypeptide chains. Each dipeptide was composed of one of the following types of amino acids: leucine (nonpolar), serine (polar, hydroxyl group), glutamine (polar, amino group), and lysine (basic). The diglycine and hexaglycine molecules were simulated in the absence and presence of TMAO.

As a starting point, the fully extended peptide was centered in a rectangular water box comprised of TIP3P water molecules, and all of the water molecules within 2.2 Å of the peptide molecule were deleted. The dimension of the water box was set to 10 Å, which is more than the length of the peptide.

We performed ten independent simulations where the TMAO positions were set by randomly replacing the TIP3P water molecules such that the concentration was 1.0 M, a value that is typically used in transfer experiments. For each independent trajectory, 100 initial configurations with different placements of the TMAO were generated, followed by 10 steps of steepest-descent and 25 steps of Adopted-Basis Newton-Raphson minimization with harmonic constraints on the peptide, and only the lowest energy configuration was used for simulations. Therefore, each independent simulation started from a unique, energy-minimized, random distribution of TMAO. We equilibrated the system by removing all harmonic constraints, applying 2,000 steps of conjugate gradient minimization, and performing 50 ps of NVT MD simulations using a 2 fs timestep. We then performed 10 ns production runs NPT simulations of each di- and hexa-peptide at 298 K using the CHARMM force field. All analyses were performed for production run.

For the A$\beta_{16-22}$ (KLVFFAE) monomer peptide simulations, the same protocol was used as with the di- and hexa-peptides except that the dimensions of the water box were cubic with each side of length 40 Å. The equilibration time period was increased to 5 ns, and the production run time length was increased to 100 ns at 300 K, of which the last 60 ns was used for analyses. Four sets of A$\beta_{16-22}$ monomer simulations were performed at TMAO concentrations of 0 M, 1.0 M, 2.5 M, and 5.0 M.

**Results and Discussion**

*TMAO Preferentially Interacts with Peptide Backbone and Basic Dipeptide Sidechains:*

For the simulations of the leucine dipeptide (Fig. 2a), the radial distribution function, g(r), between $O_T$ and the peptide backbone nitrogen (N), which is the only hydrogen bond donor, resulted in a peak with g(r≈3Å) ~ 1.5. At the typical hydrogen bond distance, r ≈ 3Å, the most probable value of the angle between N, the amide hydrogen, and $O_T$ is 150° (Fig. 3a). The g(r) between the TMAO methyl carbon ($C_T$) and the carbon atom of the terminal methyl group in the leucine sidechain has a modest peak where g(r) ~ 1.0 at, r ≈ 4Å, which shows that the dispersion interactions with the sidechain are negligible (Fig.

2a). Since leucine is the most hydrophobic residue, the interaction between TMAO and any other hydrophobic sidechain must be less favorable in comparison to hydrogen bonding with the backbone. Serine and glutamine dipeptides (Fig. 2b,c), which have hydroxyl and amino hydrogen bond donors, respectively, in the sidechains, also resulted in g(r) ~ 1.5 for the backbone nitrogen, but g(r) ~ 1.0 for the sidechain oxygen ($O_\gamma$ in serine) and nitrogen atoms ($N_\epsilon$ in glutamine). Interestingly, there is a greater preference for TMAO to hydrogen bond with the backbone nitrogen than the sidechain hydrogen bond donors, including the amino nitrogen of glutamine (Figs. 3b and 3c).

The g(r≈3Å) ~ 1.5 peak is independent of the polarity of the sidechains (Fig. 2). In lysine dipeptide, however, there is also a pronounced peak (g(r≈3Å) ~ 2.5) between $O_T$ and the charged sidechain amino nitrogen ($N_\zeta$) without compromising the hydrogen bond formation with amide nitrogen (Figs. 2d and 3d). Clearly, it is possible for charged sidechain hydrogen bond donors to have significant interactions with TMAO. We expect similar results for TMAO interactions with other basic amino acids, but their relative abundance in proteins suggests that the overall significance of TMAO-sidechain interactions may not be significant. Interestingly, the radial distribution functions in Fig. 2 also show that the size of the sidechain does not affect the extent of interactions with the backbone nitrogen atom, as long as the backbone is solvent-exposed.

The preferential interaction of TMAO with the backbone hydrogen bond donor over the uncharged sidechain hydrogen bond donor can be understood in structural terms. The peptide backbone forms a resonance interaction between the nitrogen atom and the carbonyl group. Thus, the peptide bond not only has a partial double bond character, it also leaves the nitrogen with a partial positive charge (and the carbonyl oxygen with a partial negative charge)[22]. Therefore, TMAO would form more favorable hydrogen bond interactions with the partially charged backbone nitrogen than an uncharged sidechain hydrogen bond acceptor and an even greater favorable interaction with fully charged sidechain amino nitrogen, as explicitly shown for lysine. Of course, asparagine and glutamine sidechain amides can also participate in resonance stabilization such that the amide nitrogen has a partial charge, but the electronegativity of its neighbors makes its overall partial charge less than that of the peptide amide nitrogen.

*TMAO-peptide group interactions depend on the length of the polypeptide:*

The dipeptides are not long enough to form intramolecular interactions. Thus, it is not possible to investigate the propensity to form TMAO-induced secondary strucure formation. Also **in a recent paper[21] by Neuweiler et al., it was found that the backbone hydrogen bonds are primarily responsible for the collapse of the peptide chains.** To understand osmolyte-induced changes in the collapse and secondary structural elements, and the dependence of polypeptide length on TMAO interactions we simulated hexaglycine ($G_6$) for which the conformational space can be fully explored on a nanosecond timescale. Although $G_6$ is unlikely to form a well-ordered secondary structure because the enthalpically favorable hydrogen bonds cannot overcome the entropy loss, TMAO can alter the population of the most probable ($\phi,\psi$) angles that are adopted in water. The polyglycine chain is also an excellent model system for the study of backbone dynamics in pure water and in TMAO solutions due to the absence of sidechain moieties[16]. The backbone dihedral angles of $G_6$ populate four sets of conformations that correspond to the left and right handed α-helices and PPII β-sheet like conformations in water (Fig. 4a). The addition of 1 M TMAO shifts the Ramachandran folding free energy profiles by expanding the region corresponding to the α–helical basin (Fig. 4b). The distribution of the radius of gyration ($R_g$) over all ten 10 ns trajectories shows a clear increase in the relative population of structures with $R_g$ close to that of an ideal α-helical $G_6$ (Fig. 4c,d). Interestingly, the $<R_g>$ is similar to the value obtained by Shortle and coworkers from their NMR and SAXS measurements, which show a value intermediate between ideal α-helices and PPII β-sheets[23].

The radial distribution functions involving the atomic interactions of water are almost quantitatively identical in pure water and 1 M TMAO (see Fig. 5), even in the presence of hexaglycine. These results show that the structural changes observed in $G_6$ have to be related to the interactions (or depletion) of TMAO with the polypeptide chain.  Since α-helical $G_6$ conformations can have up to two intramolecular hydrogen bonds of the backbone nitrogen (with the backbone carbonyl oxygen that is separated by four amino acids earlier in sequence), we calculated the g(r) between the N and $O_T$.  The g(r) for $G_6$ is much lower than diglycine $(G_2)$ (Fig. 6a,b), which we use as a control because it is too short to form intramolecular hydrogen bonds.  The enhancement of the α-helical basin in 1 M TMAO in $G_6$ is due to its depletion from around the peptide backbone, which is in accord with previous studies[2].  Thus, the formation of α-helical structure disfavors backbone hydrogen bond formation with TMAO, but the backbone amide nitrogen still favors hydrogen bonding with the carbonyl oxygen because of its larger dipole moment and proximity as compared to the TMAO oxygen.  The differences in g(r) between the N and $O_T$ in $G_2$ and $G_6$ highlight the role of chain length in TMAO-peptide interaction.  For $G_2$, the amide nitrogen is accessible to $O_T$, which is consistent with simulations of cyclic $G_2$[5].

*TMAO is a Nano-Crowding Agent:*

Observations of ordered secondatry structure in 1 M TMAO in $G_6$ can be rationalized using the depletion theory used to predict stability changes in the folded states in a crowded environment. Depletion of TMAO around $G_6$ essentially induces an osmotic pressure[24] on $G_6$, **which results in chain compaction as was previously observed in MD simulations of the longer $G_{15}$ in TMAO[25]**.  Thus, the polypeptide is forced to adopt conformations that maximize intramolecular interactions.  In the $G_6$ case this results in an increase in the population of α-helical structure (Fig. 4b).  The exclusion of TMAO from $G_6$ is vividly illustrated using a number of pair functions involving water, TMAO, and the amide nitrogen (Figs. 7a and 7b).  Thus, in 1 M TMAO, $G_6$ is localized in a region that is devoid of both water and TMAO.  Such a mechanism is exactly the one invoked to quantitively predict the native state stabilization in a crowded environment due to volume excluded to the protein by the crowding particles!

Two remarks of caution are in order. (1) For small peptides such as $G_2$, the amide nitrogen interacts favorably with $O_w$ and $O_T$ (Fig. 7c). Thus, polypeptide chain length is important in observing TMAO-induced structure formation. (2) More importantly, it is known from crowding theory[11,12] that the nature of the structures adopted depends on $q = \left\langle R_g \right\rangle \big/ R_c$ where $R_c$ is the size of the crowding particles. In our study of $G_6$ in TMAO, $<R_g> = 5.01$ Å and $R_c = 1.32$ Å, resulting in q ≈ 3.77. It is unclear whether depletion theory also holds if q and sequence are varied.

*$A\beta_{16-22}$ Peptide Becomes Compact and $\alpha$-Helical with Increasing TMAO Concentration:*

In order to assess if the depletion mechanism leading to a shift in the population towards $\alpha$-helical structure is general we performed additional simulations of TMAO-induced changes in $A\beta_{16-22}$ (KLVFFAE) monomer peptide, which aggregates to form antiparallel fibrils[21] at high peptide concentration. We had shown earlier that $A\beta_{16-22}$ monomer is a random coil largely devoid of secondary structure. The population of $\alpha$-helical structure is less than about 1%. If TMAO acts as a nano-crowder then we expect that $A\beta_{16-22}$ woud be localized in a region devoid of TMAO. Under these conitions $A\beta_{16-22}$ is expected to adopt $\alpha$-helical conformation to maximize intramolecular interactions[26]. In order to test the applicability of depletion-induced structure formation we performed simulations of $A\beta_{16-22}$ in various TMAO concentrations.

In addition to being a biologically relevant system, the intrinsically disordered $A\beta_{16-22}$ peptide is a very good model system to study the role of TMAO on conformational fluctuations of peptides since its sequence consists of charged residues, a positive lysine (K) and a negative glutamic acid (E), that cap the ends of a short stretch of hydrophobic residues. The probability distribution of the radius of gyration, $P(R_g)$, of the $A\beta_{16-22}$ peptide shows that it becomes more compact with increasing TMAO concentration (Fig. 8). The values of mean $R_g$, $<R_g>$ ($= \int P(R_g) dR_g$) for the TMAO concentrations, [TMAO] = 0 M,

1.0 M, 2.5 M, and 5 M, are 6.9 Å, 6.7 Å, 5.9 Å, and 5.7 Å, respectively. Thus, there is a 17% reduction in $<R_g>$ as [TMAO] is changed from 0 to 5 M.

We further probed the structural changes by calculating the Ramachandran free energy profiles of the peptide. In the absence of TMAO Aβ$_{16-22}$ fluctuates among a number of distinct structures. Fig. 9a shows that Aβ$_{16-22}$ the basins with ($\phi,\psi$) angles that correspond to β-sheets, as well as left- and right-handed α-helices are populated (Fig. 9a), as would be expected from an intrinsically disordered peptide that is basically a random coil. At a modest concentration of TMAO (i.e., [TMAO] = 1 M), the β-sheet basins disappear (Fig. 9b). Remarkably, for [TMAO] greater than 2.5 M only the right-handed α-helices remain (Fig. 9c,d). Thus, TMAO-induces a transition between a predominantly random coil state to α-helical structure. Considering the small size of Aβ$_{16-22}$ the transition is relatively sharp (Fig. 9e). Secondary structure analysis[26] is performed using the dihedral angles φ and ψ. For a β-strand, φ and ψ satisfy 150° ≤ φ ≤ 90° and 90° ≤ ψ ≤ 150°, and for a α-helix 80° ≤ φ ≤ 48° and 59° ≤ ψ ≤ 27°. A conformation is in a β-strand (α-helix) if (i) at least 2 consecutive residues adopt strand (helix) configuration, and (ii) no 2 consecutive residues are in helix (β-strand) state. The fraction α-helical (β-strand) secondary structure content in a conformation for the 5 internal residues ($^2$LVFFA$^6$) is defined as, $f_{ss} = \dfrac{1}{5}\sum_{i=1}^{5}\delta_{i,\beta}$, and $\delta_{i,\beta}$=1 if residue $i$ satisfies conditions for α-helix (β-strand) else it is 0. The equilibrium fraction secondary structure is obtained by averaging atleast 60ns of simulation data for each TMAO concentration. Results similar to the random coil to α-helix transition on adding TMAO are experimentally observed where TMAO induces helical formation in alanine peptides[27].

The tendency of Aβ$_{16-22}$ to form ordered α-helical structures have implications for oligomer formation. The formation of stable helical structure could preclude amyloid formation, which requires β-structures as seeds. It is interesting to constrast TMAO-induced structure formation to the effects of urea on Aβ$_{16-22}$. Molecular dynamics showed that in urea Aβ$_{16-22}$ monomer is extended and forms β-strands[28]. The contrasting behavior could have implications for aggregation in mixed cosolvents containing urea and TMAO.

*TMAO Interacts with Aβ₁₆₋₂₂ Backbone and Lysine Sidechain:*

To determine the molecular interactions that induce the helical formation of the Aβ$_{16-22}$ (KLVFFAE) peptide, we calculated the radial distribution function, g(r), between TMAO and the Aβ$_{16-22}$ peptide. There is a stronger preference for terminal (K16, L17, A21, and E22) backbone amide N with TMAO (Fig. 10a) compared to those in the interior (V18, F19, and F20) (Fig. 10b), which may be a reflection of the bulky phenylalanine that effectively excludes interactions with the peptide backbone. The TMAO interactions with the backbones of these residues are more pronounced for polar interactions with the amide N (Fig. 10c,d). **The residence of TMAO near the backbone atoms is approximately 55 picoseconds which is approximately twice of that of water. The residence time is defined as the time during which any of the TMAO or water atoms are within 4Å of any of the backbone atoms.** Hydrophobic interactions with TMAO are modestly significant for the sidechains (Fig. 10e) and nonexistent in our simulations for the backbone (Fig. 10d). The interactions between TMAO and the terminal positively charged lysine sidechain, however, is pronounced (Fig. 10f), even more than interactions with the terminal backbone amide N (Fig. 10a). The affinity of TMAO to negatively charged sidechains is minimal (Fig. 10f).

**Conclusions**

Using all-atom MD simulations of a number model peptide constructs and Aβ$_{16-22}$ monomer in TMAO solution, we dissected the molecular mechanism of how TMAO stabilizes the native basin of proteins. By preferentially hydrogen bonding to the backbone nitrogen of the solvent exposed peptides, TMAO acts as a nano "crowder" that limits the degrees of freedom of the unfolded state and entropically destabilizes it. When the backbone nitrogen forms secondary structure, it is no longer available to hydrogen bond with TMAO resulting in the depletion from the vicinity of the protein, which in turn results in native state stabilization. Comparisons between $G_2$ and $G_6$ shows that polypeptide length is a relevant factor in determining the energetic balance between collapsed and extended structures. If the polypeptide chain exceeds a critical size, it is likely that aqueous TMAO would be a "poor" solvent for generic proteins, which would promote collapse and structure formation as demonstrated for $A\beta_{16-22}$ peptide. Finally, our work also shows that in sequences that contain charged residues (e.g., intrinsically disordered proteins or fragments of $A\beta$ peptides such as $A\beta_{16-22}$), the interactions between TMAO and positively charged sidechains are significant. However, for generic proteins, TMAO is expelled from the surface. In this sense TMAO behaves as a nano-crowding particle, thus stabilizing proteins by the entropic stablization mechanism[10].

**Acknowledgement.** This work was supported in part by a grant from the National Science Foundation (CHE 09-10433). SSC was supported by a Ruth L. Kirschstein National Research Service Award from the National Institutes of Health.

**Supporting Information Available**. Radial distribution functions between amide N and H in dipeptide constructs and oxygens of water and TMAO. Pair functions between amide N and H in glycine constructs and oxygens of water and TMAO. Distribution of hydrogen bond angles formed between dipeptide and hexapeptide constructs of glycine and TMAO.

**Figure Captions:**

**Figure 1.** Structure of TMAO. The partial charges of the N-oxide group and the distribution of the dipole moment are identified. The values of partial charges, $\delta^+$ and $\delta^-$, are 0.44 and -0.65, respectively.

**Figure 2.** Radial distribution functions between atomic centers of TMAO and dipeptides. (a) Leucine, (b) Serine, (c) Glutamine, and (d) Lysine. The amino acid chemical structures are at the top, and the corresponding radial distribution functions are below. The corresponding distribution of angles formed by hydrogen bonds formed by the peptide backbone amide N and H with the TMAO oxygen ($O_T$) are shown in Fig. 3, and the pair functions involving other interaction sites are shown in Fig. S1.

**Figure 3.** Distribution of angles formed by hydrogen bonds between the peptide backbone amide N and H with the TMAO oxygen ($O_T$) for the dipeptide constructs of a) leucine, b) serine, c) glutamine, and d) lysine, and e) glycine, as well as the hexaglycine construct. The angles exceeding 150°, which are characteristic of a perfect hydrogen bond, are most probable. Only interactions for which the distance between the backbone amide and TMAO is less than 3.5 Å, which corresponds to the first solvation shell, are considered. Thus, at the distance when g(r) has a first peak in all dipeptides, $O_T$ forms a hydrogen bond with the amide proton.  See Fig. S2 for the corresponding hydrogen bond distribution of angles for the dipeptide and hexapeptide constructs of glycine.

**Figure 4.** Conformations adopted by hexaglycine in the absence (a,c) and presence of 1M TMAO (b,d). The Ramachandran free energy profiles are shown with the four major basins labeled (a,b).  The normalized histograms of the radius of gyration, $R_g$, are shown with the $R_g$ of ideal $\alpha$-helices and PPII $\beta$-sheets labeled.

**Figure 5.** Comparison of the radial distribution functions involving water interaction sites in pure water (0M) and 1M TMAO solution for hexaglycine simulations. The inter-water O-O, O-H, and H-H radial distribution functions, g(r), are shown. The structure of water is not significantly perturbed in TMAO solution containing $G_6$.

**Figure 6.** Pair functions between TMAO and (a) diglycine and (b) hexaglycine.

**Figure 7.** Pair functions between various amide backbone atoms from (a,b) hexaglycine and (c) diglycine  constructs and atomic centers on water and TMAO. The top row consists of radial distribution functions, g(r), between the backbone amide nitrogen (N) and the oxygens of water ($O_w$) and TMAO ($O_T$).  The bottom row is the same except with backbone amide hydrogen (H).  In the presence of TMAO, the strength of the hydrogen bond involving the amide N is suppressed.  The decrease is dramatic when the results for $G_2$ and $G_6$ are compared.  Interestingly, the amide N in $G_6$ does not even form hydrogen bonds with $O_w$ when TMAO is present.  Thus, both the solvent and the solute (TMAO) are depleted from the surface of $G_6$. This effect is essentially similar to the entropic stabilization in the excluded volume dominated crowding agents.

**Figure 8.** Probability distribution of $R_g$ of A$\beta_{16-22}$ in various TMAO concentrations.

**Figure 9.** Ramachandran free energy profiles of A$\beta_{16-22}$ at TMAO concentrations of (a) 0 M, (b) 1 M, (c) 2.5 M, (d) 5 M (e) Fraction of secondary structure content, $f_{ss}$, as a function of TMAO concentration. The lines are a guide to the data points.

**Figure 10.** Radial distribution functions between atomic centers of TMAO and A$\beta_{16-22}$ for 1M TMAO concentration. (a) The TMAO oxygen ($O_T$) interactions with the backbone amide N of the termini residues of the A$\beta_{16-22}$ peptide. (b) The $O_T$ interactions with backbone amide N of residues in the interior of the A$\beta_{16-22}$ peptide.  The only significant TMAO interactions with the backbone amide N are observed for those in the N-terminal residues.  (c) A comparison of the total backbone polar amide N with $O_T$ vs. hydrophobic $C_\alpha$ interactions with the TMAO carbon $C_T$, as well as (d) the individual per residue hydrophobic $C_\alpha$ interactions. There are no significant hydrophobic backbone $C_\alpha$ interactions with TMAO. (e) Hydrophobic sidechain interactions with $C_T$ show that these interactions can be modestly significant.  (f) The polar sidechain interactions with $O_T$ show a clear and significant preference of TMAO for the positively charged lysine sidechain.

SYNOPSIS TOC

The osmolyte TMAO, which can be thought of a chemical chaperone, accumulates in the cell in response to osmotic stress. The role of TMAO in the stabilization of proteins has been widely studied experimentally, but the precise molecular mechanism is not clear. We performed all-atom simulations of several model peptides in order to describe in atomic detail the origin of TMAO-induced stabilization of proteins. We find that TMAO preferentially hydrogen bonds with the dipeptide backbone, but generally not with nonpolar or polar sidechains unless they are positively charged in the unfolded state. In hexaglycine, however, TMAO is depleted from the protein backbone. To verify our hypothesis, we performed simulations of the intrinsically disordered A$\beta_{16-22}$ (KLVFFAE) monomer, in various concentrations of TMAO, and we observe unambiguous $\alpha$-helical formation at high concentrations. We observe with a clear preference of TMAO for polar interactions with the backbone amide N or positively charged sidechains. We argue that the origin of these interactions is the TMAO dipole moment that favors interactions with partially or fully positively charged moieties of amino acids in proteins. Our simulations also highlight the potential similarities between the action of TMAO and entropic stabilization of proteins in a crowded environment in which the interactions between protein and the crowding agents are repulsive.

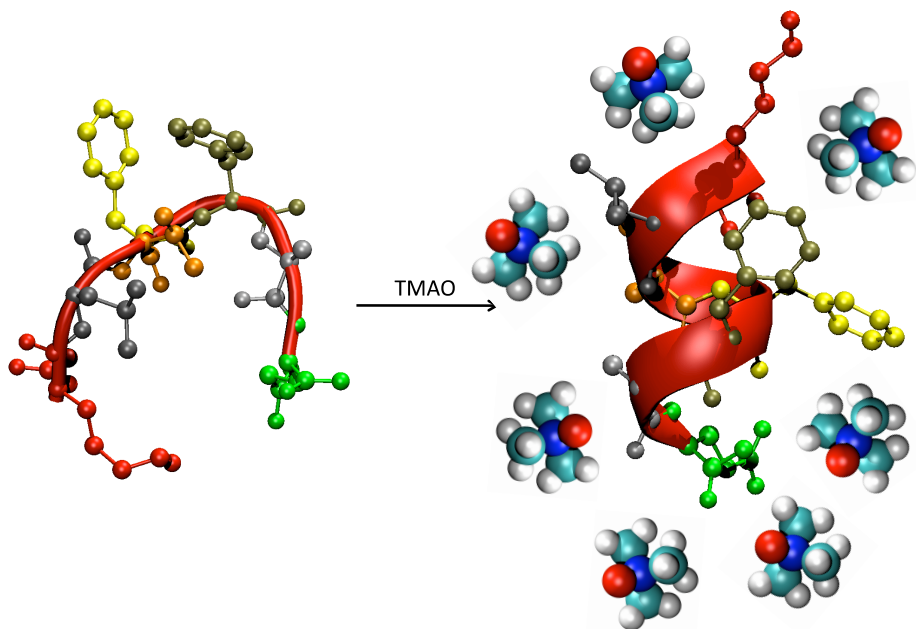

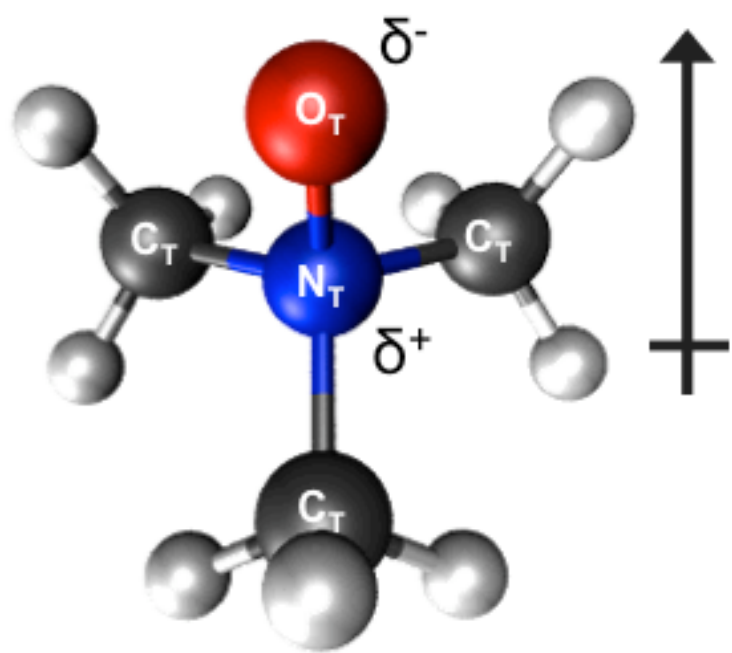

δ⁻

δ⁺

**Figure 1**

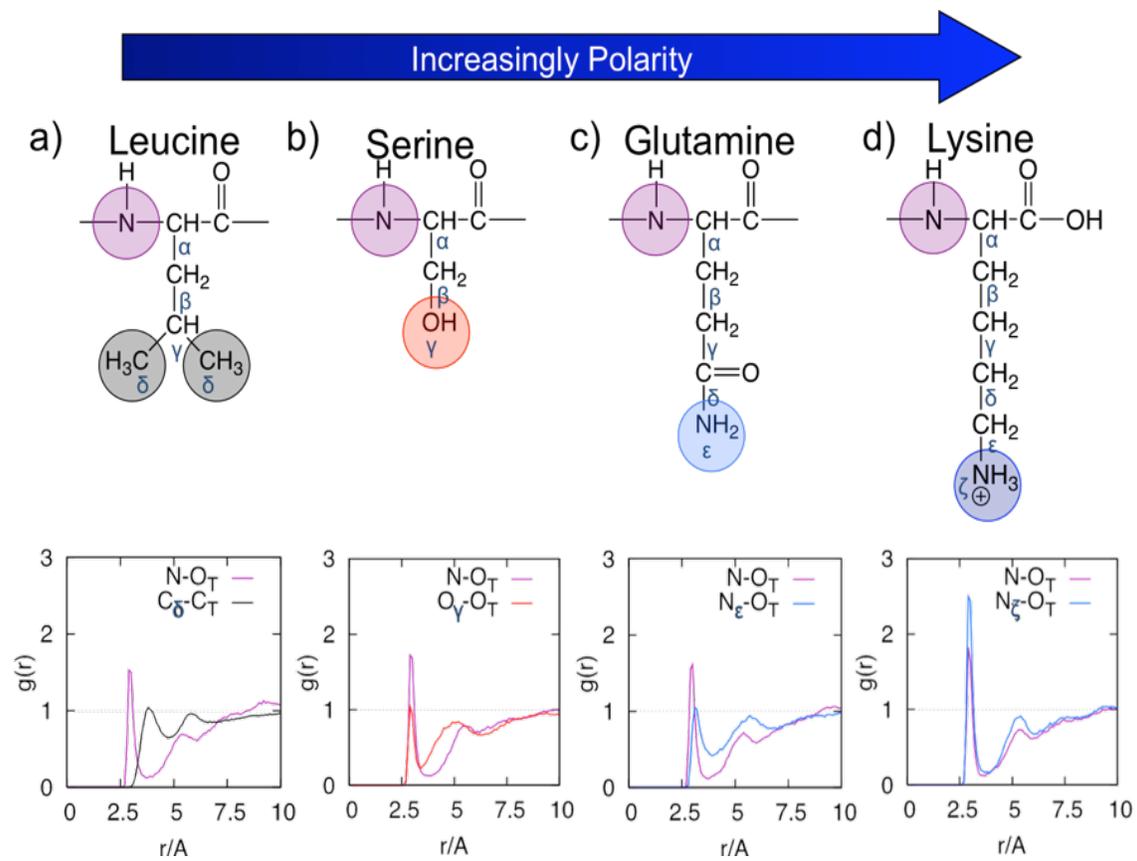

**Figure 2**

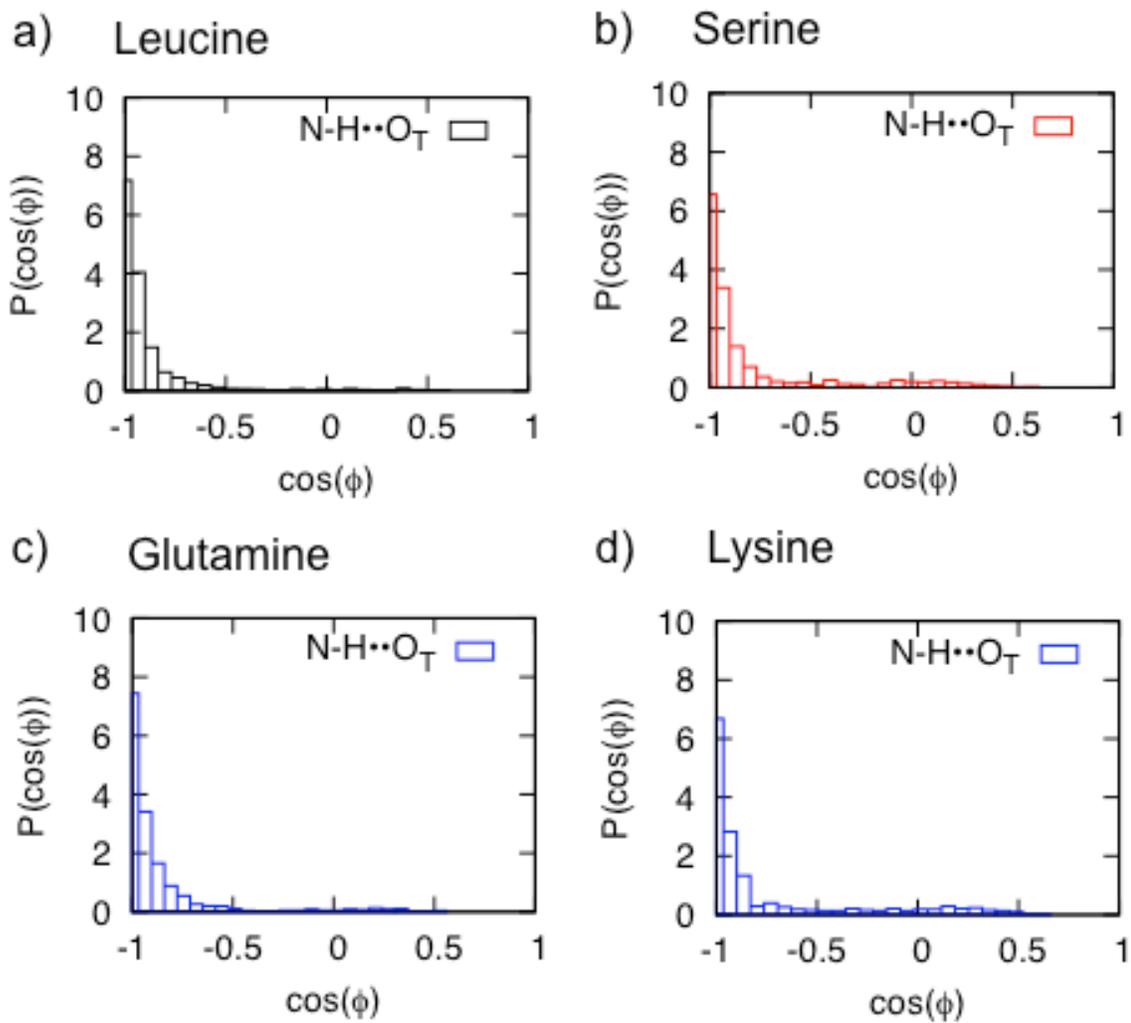

**Figure 3**

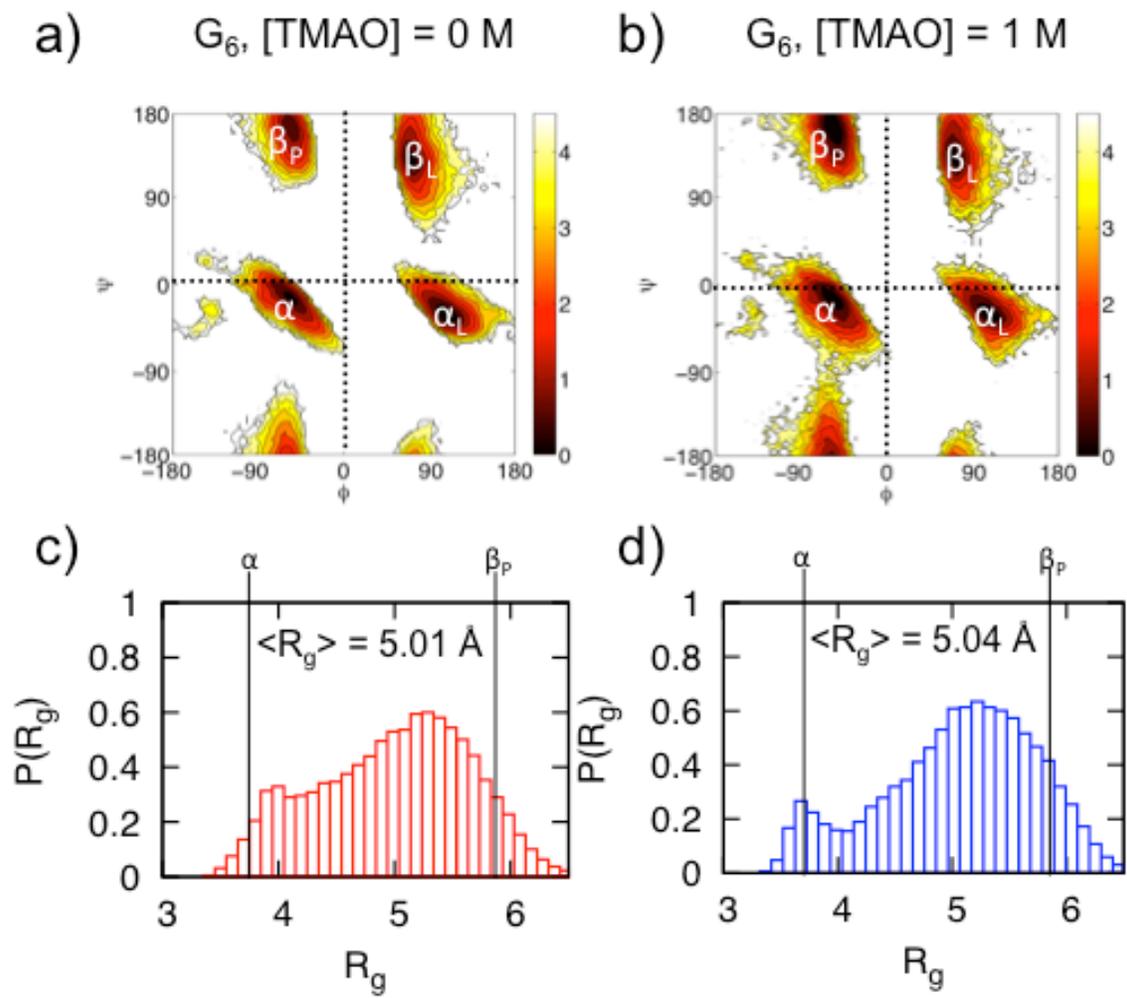

**Figure 4**

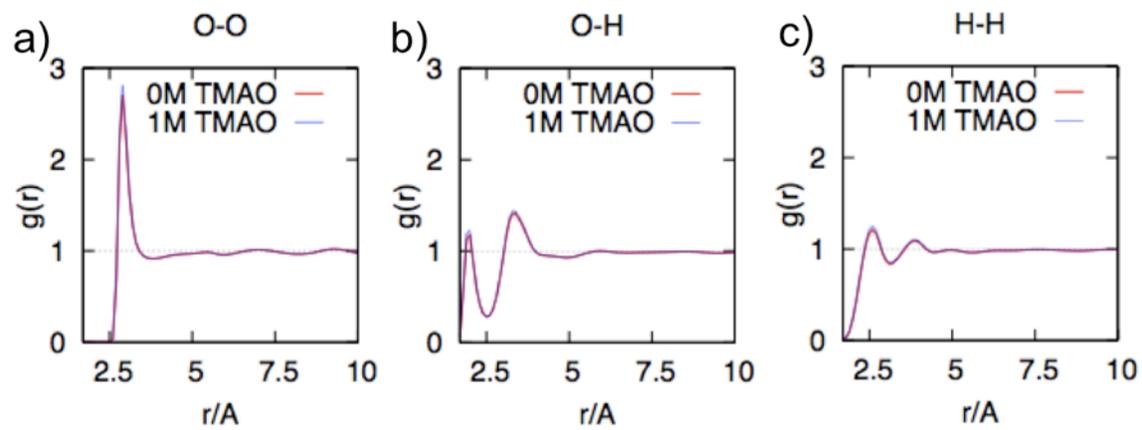

**Figure 5**

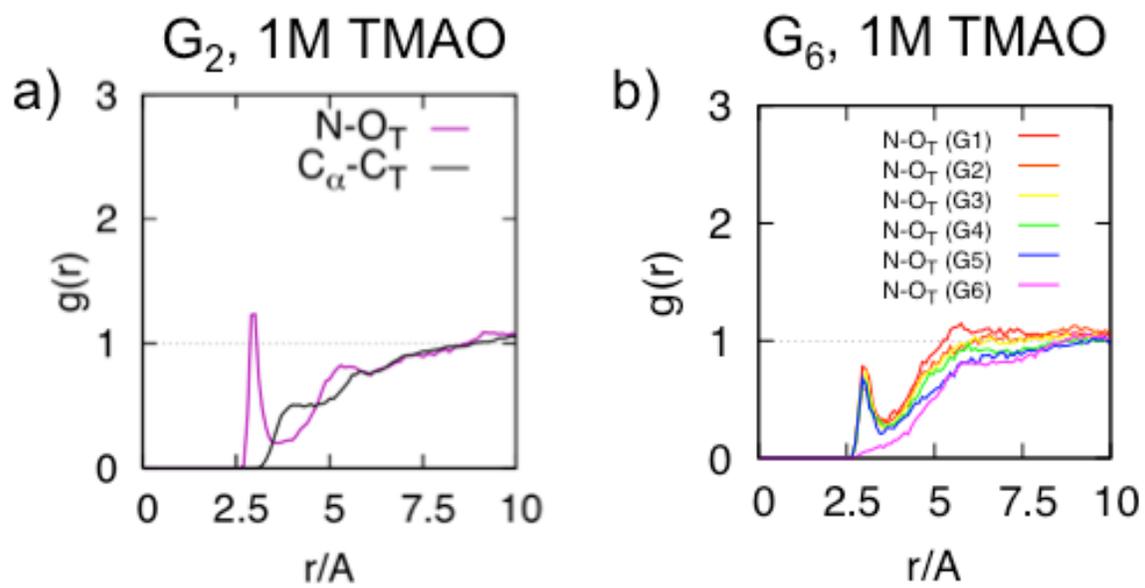

**Figure 6**

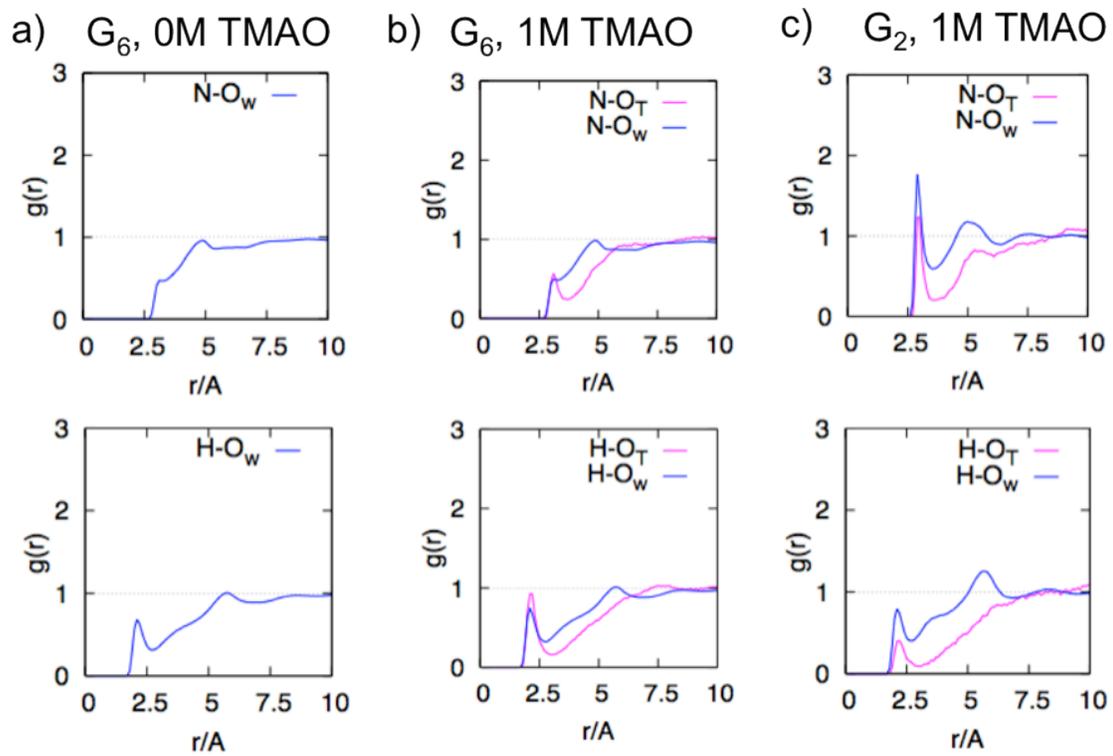

**Figure 7**

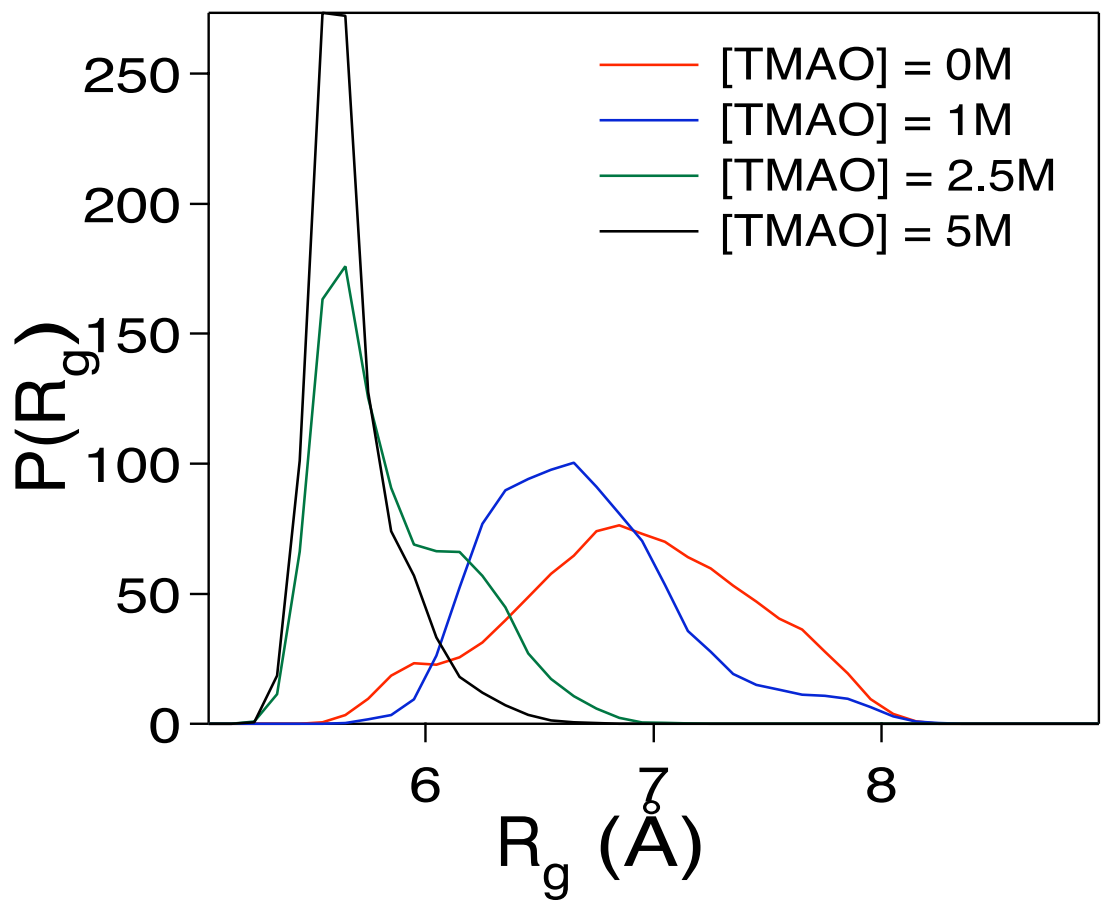

Figure 8

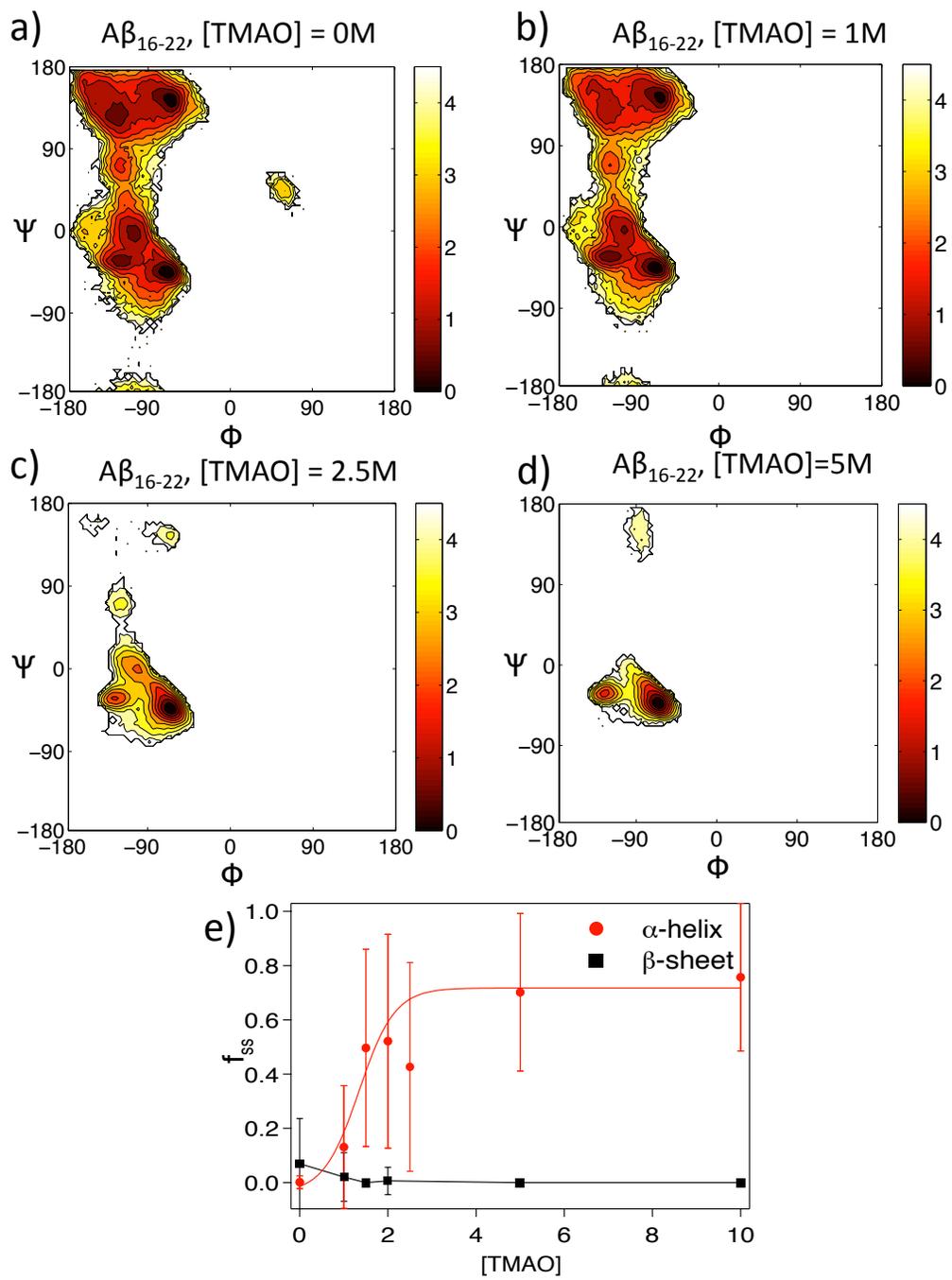

**Figure 9**

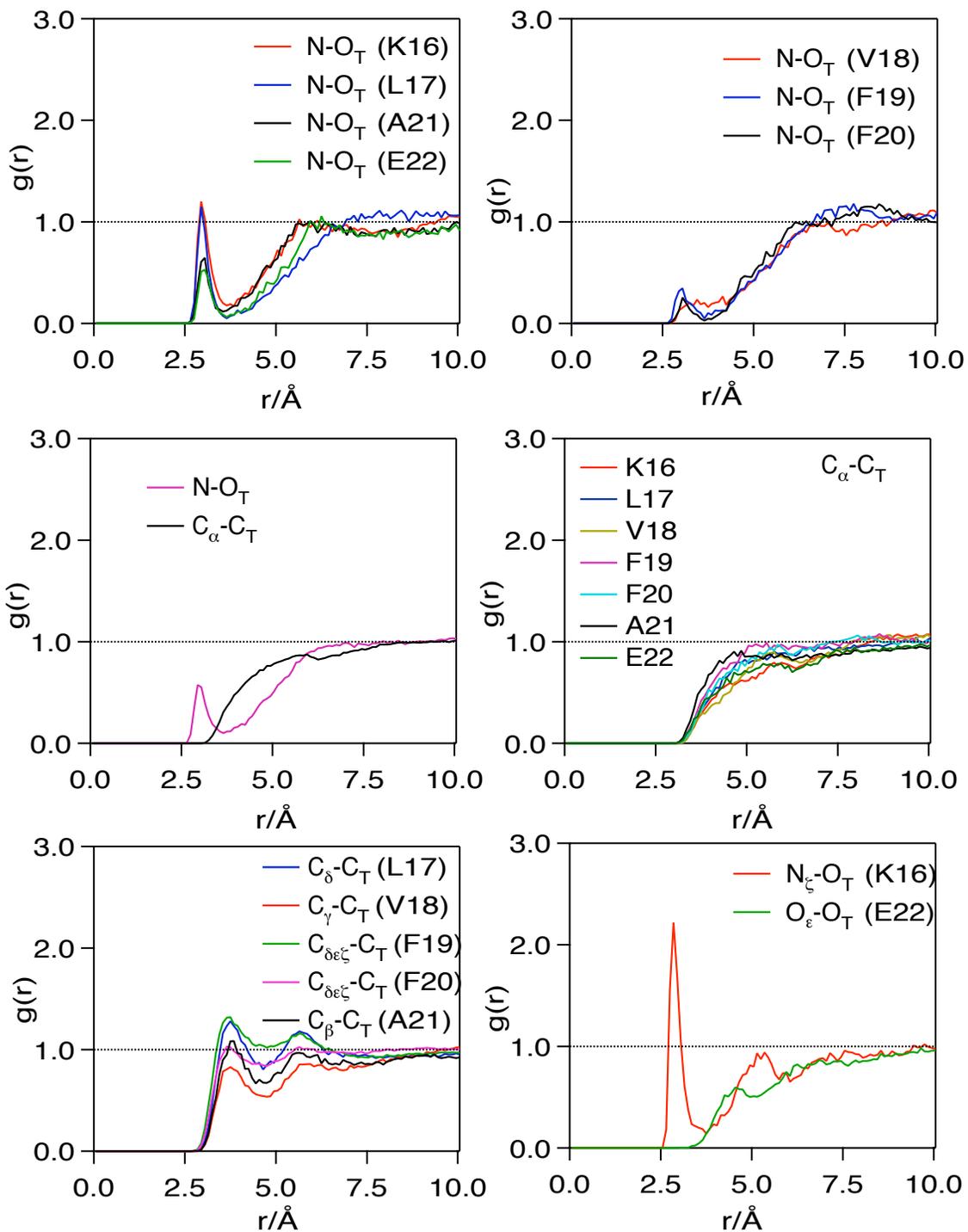

**Figure 10**

## SUPPLEMENTARY INFORMATION

FIGURES

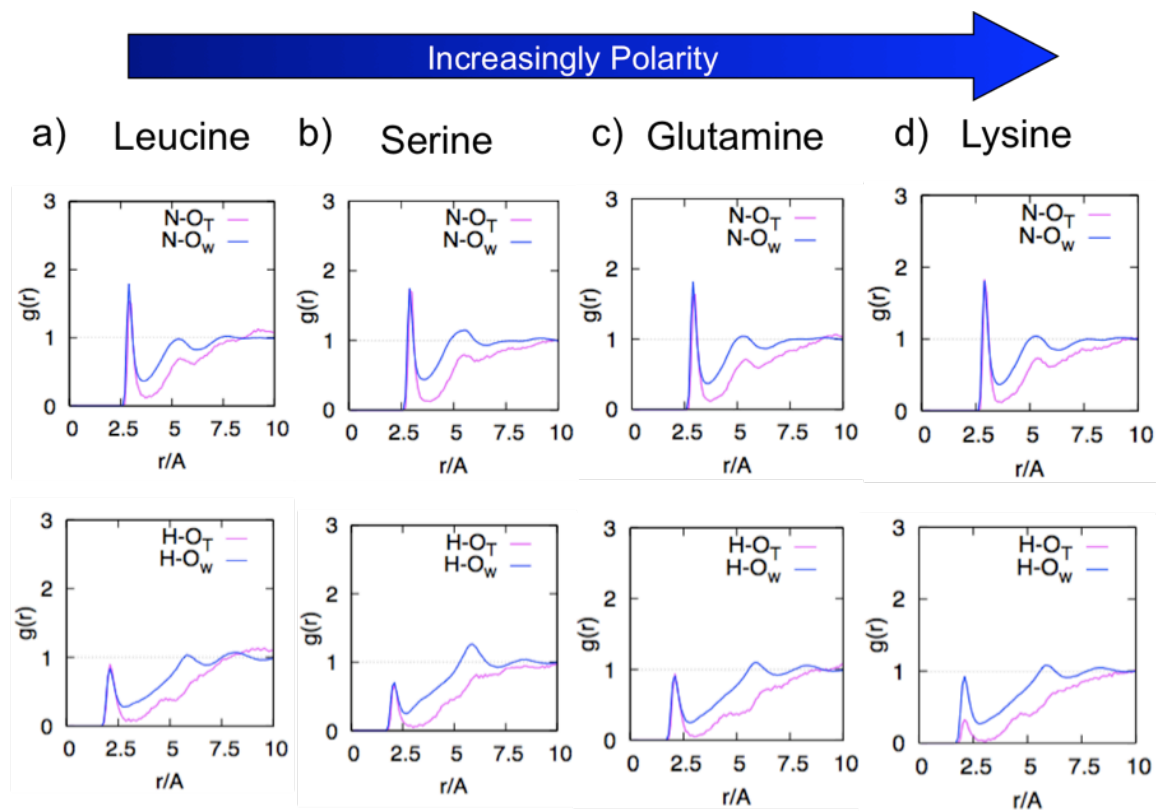

**Figure S1.** Radial distribution functions (g(r)'s) between various dipeptide amide backbone atoms and sites on water and TMAO. The top row shows g(r)'s between the backbone amide nitrogen (N) and the oxygens of water ($O_w$) and TMAO ($O_T$). The bottom row is the same except with backbone amide hydrogen (H). The pair function between N and $O_T$ (see also Fig. 2 in the main text) is shown for comparison. The pair functions between $O_w$ and $O_T$ and the sites in the backbone (N and H) are virtually independent of the polarity of the sidechains.

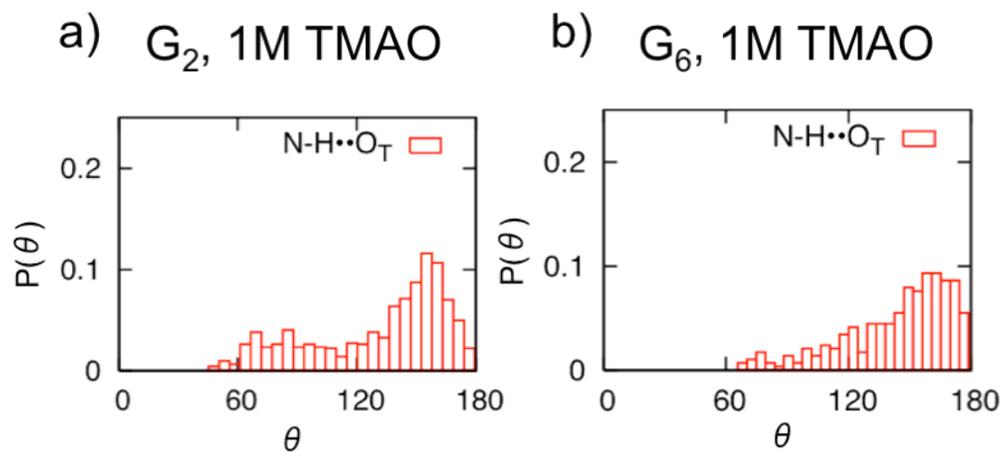

**Figure S2.** Distribution of angles formed by hydrogen bonds between the peptide backbone amide N and H with the TMAO oxygen ($O_T$) for the a) dipeptides and b) hexaglycine constructs glycine. The angles exceeding 150°, which are characteristic of a perfect hydrogen bond, are most probable in both cases, but significantly less prominently for diglycine. Only interactions for which the distance between the backbone amide and TMAO is less than 3.5 Å, which corresponds to the first solvation shell, are considered.